"Low Mass Higgs Boson Consistent with Precision Experiments: A Consequence of Large Top-Yukawa Coupling in Condensate Models"

Bipin R. Desai  and  Alexander R. Vaucher

Department of Physics, University of California
Riverside, California 92521, USA

Abstract

It is shown, using dispersion relations techniques for bound states, that the presence of a large top-Yukawa coupling lowers the Higgs mass from the condensate-model value of twice the top mass ($\approx 350$ Gev) to $100-200$ Gev consistent with the $Z^0$ width precision measurements. The coupling is found to be $\approx 3.7$ at the top-mass, much larger than the Standard Model value $\approx 1$. It corresponds to a compositeness scale $\approx 1.4$ Tev, which is consistent with top-color models, and implies quite different scales for fermion mass generation and electroweak symmetry breaking. A second scalar state around 1 Tev also emerges as a solution in combination with the low mass Higgs.

# Introduction

It has been suggested recently, particularly in the framework of the top color models, that the fermion mass generation and electroweak symmetry breaking, which have the same origin in the standard model with a single Higgs doublet, may actually have different origins [1], [2], [3]. In a composite model such as the top-condensation model of Nambu, [4] and Bardeen et al [5], this allows the composite scale to be in the Tev range rather than at the GUT scale [5], and, therefore, allows the top Yukawa coupling $\lambda_t$ to be large ($\lambda_t = \frac{\sqrt{2}\,m}{v}, v << 246\,Gev$). The origin of the electroweak symmetry breaking is then assumed to lie elsewhere.

The top Yukawa coupling evolves according to the renormalization group equations as:

$$\mu^2 \frac{d\lambda_t}{d\mu^2} = b\lambda_t^3 \qquad (1)$$

where $\mu$ is the energy at which $\lambda_t$ is measured, and $b$ is a positive constant, its magnitude depending on the assumed higher symmetry. The Yukawa coupling of the other quarks is ignored as well as the gauge couplings since $\lambda_t$ is expected to dominate. We note from (1) and from ref. [5] that, with the standard model restriction $\lambda_t \approx 1$ lifted, a large coupling at the top mass could lead to a rapidly diverging $\lambda_t$ for large $\mu$, a fact that will have an important bearing as we will see below.

In top condensation models the Higgs boson is generated from a four-fermi interaction at a high scale by summing the fermion bubbles in the large $N_c$ limit. The $t\bar{t}$ amplitude in $J^P = O^+$ channel (the s-channel) develops, as a result of the summation, a pole at $s = 4m^2$, where $m$ is the mass of the top. The Higgs boson, therefore, will have a mass $m_H = 2m$, which is about 350 Gev. However, precision measurements on the $Z^0$-width appear to indicate the Higgs mass to be lower, between 100 and 200 Gev [6], [7].

It is interesting to examine whether , because of the large magnitude of the top-Yukawa coupling, the condensate value $2m$ could change, in particular, whether it would go down. At the same time it is also interesting to examine whether a strong top-Yukawa coupling could dynamically generate further scalar states which have been speculated in several models.

The indirect measurements on the Higgs mass referred to earlier assume the standard model to be correct [6], [7]. Our use of the condensate models and top color models to explain those results will not be incompatible with those assumptions because they reproduce the standard model at low energies.

We will try to solve the bound state problem using the ladder approximation of iterating the lowest order Higgs pole diagram. As for the strength of the top-Yukawa coupling, $\lambda_t$, even though estimates for its magnitude are obtained in terms of the cut-off partameter [8] we will keep it as a free parameter to be varied in order to produce the bound states at appropriate energies.

## **Bound State Calculation**

Before we embark on our calculation we will first construct the $t\bar{t}$ amplitude which has the same quantum numbers in the s-channel as the Higgs boson (i.e. $J^P = O^+$). For this we use the Jacob and Wick [9] helicity formalism for states, $|\lambda_1, \lambda_2\rangle$, with helicities $\lambda_1 = \pm\frac{1}{2}$, $\lambda_2 = \pm\frac{1}{2}$, with the scattering amplitude $ab \to cd$ given by

$$T_{cd,ab}(E) = \frac{1}{p}\sum_j \left(j+\frac{1}{2}\right)\langle \lambda_c \lambda_d |T_j(E)|\lambda_a \lambda_b\rangle e^{i(\lambda-\mu)\phi} d^j_{\lambda\mu}(\theta)$$

where $\lambda = \lambda_a - \lambda_b$, $\mu = \lambda_c - \lambda_d$. For $J^P = 0^+$ in the s-channel, the appropriate linear combination of the states is

$$|+\rangle = \frac{|++\rangle + |--\rangle}{\sqrt{2}}$$

where $+$ and $-$ indicate $+\frac{1}{2}$ and $-\frac{1}{2}$ respectively. For our scattering amplitude, therefore, $\lambda = 0 = \mu$ and

$$d^j_{\lambda\mu} = P_j(\cos\theta)$$

Denoting this amplitude as $T_+$, we write

$$T_+(E,\theta) = \frac{1}{p}\sum_j \left(j+\frac{1}{2}\right)\langle +|T_{+j}(E)|+\rangle P_j(\cos\theta)$$

The $j = 0$ projection $T_{+0}$ is what interests us:

$$T_{+,0}(E) = \frac{1}{2}\int_{-1}^{1} d\cos\theta \; T_{+}(E,\theta) \qquad (2)$$

The bound state problem in the ladder approximation has been studied by Lee and Sawyer in $\lambda\phi^3$ field theory [10]. They showed that it can be cast in the so called N/D formalism which has a rigorous foundation in potential theory for well-behaved potentials like the Yukawa potential [11] and which has been commonly used in S-matrix theories [12], [13]. The solution is obtained through dispersion relations by imposing the known analytic and unitarity properties of partial wave scattering amplitudes. The zeroes of D, then determine the bound state (or resonance) energies.

Ghergetta [14] has shown that the dispersion relations method of the type indicated above is better suited than the traditional methods to regulate the Nambu Jona-Lasinio model [15], particularly the top quark condensation model of Nambu [4] and Bardeen et al [5] by maintaining gauge invariance without depending on arbitrary shifts in loop momenta involved in the fermion bubble summation.

In the following we will attempt to solve our Higgs problem through this method. Typically, in this formalism, a partial wave amplitude is expressed as a ratio

$$T_l = \frac{N}{D} \qquad (3)$$

In order to discuss the dispersion relations we first define a variable $v = p^2$ ($p$ being the c.m. momentum). The partial wave amplitude, $T_l$, generally has two types of branch cuts in the complex $v$ plane. One related to $v \geq 0$ where $T_l$ becomes complex, acquiring a phase,

$$T_l = \frac{e^{i\delta} \sin\delta}{\rho}, \qquad v \geq 0 \qquad (4)$$

where $\delta$ is the phase-shift, and $\rho$ the phase space factor. The associated branch-cut (the right hand cut) is taken along $(0,\infty)$ on the real v-axis. The second type of cut exists typically whenever t-channel exchange is involved. For a particle of mass $M$, exchanged in the t-channel, the (total) scattering amplitude has the form

$$\frac{1}{t - M^2}$$

The partial wave projection, $T_l$, of the above amplitude then has a cut (the left hand cut) along $(-\infty, v_0)$ on the negative $v$ axis where $v_0 = -\frac{1}{4}M^2$. Apart from the two types of cuts, $T_l$ is real along the real $v$ axis but may have poles in $(v_0, 0)$ which correspond to bound states.

In the N/D formalism, D inherits the right hand cut, N the left, otherwise they are real along the real $v$ axis. The zeroes of D then correspond to bound states.

For $v \geq 0$, one can write $T_l$ given by (4) as

$$T_l = \frac{1}{\rho \cot \delta - i\rho}$$

Therefore comparing with (3) the discontinuity of D along the right hand cut is given by

$$\operatorname{Im} D = -\rho N \qquad 0 \leq v \leq \infty$$

similarly from (3) the N-discontinuity is given by

$$\operatorname{Im} N = D \operatorname{Im} T \qquad -\infty \leq v \leq v_0$$

The dispersion relations for N and D can now be written. Normally one would normalize the amplitude by taking $D_l \to 1$ as $v \to \infty$, i.e. write the dispersion relations as

$$D_l(v) = 1 - \frac{1}{\pi} \int_0^\infty \frac{dv'}{v' - v} \rho N_l(v') \tag{5}$$

$$N_t(v) = \frac{1}{\pi} \int_{-\infty}^{v_0} \frac{dv'}{v' - v} D_t(v') \operatorname{Im} T_t(v') \qquad (6)$$

However in our case it is essential to take account of the presence of the Higgs pole at $s = m_H^2$ (see expression (7) below). We can accomplish this by making appropriate subtractions.

The Higgs pole for $J^+ = 0^+$ amplitude, $T_{+,0}(E)$ defined in (2) in the s-channel is given by [5]

$$T_{+0} = -\frac{2\lambda_t^2 p^2}{s - m_H^2} \qquad (7)$$

In order to properly incorporate the subtraction procedure for D we note that the ladder approximation (i.e. the underlying N/D formalism) involves a sum over the powers of the coupling strength, $\lambda_t$. It should therefore reproduce (7) for small $\lambda_t$. Therefore,

$$D \approx s - m_H^2 \qquad (8)$$

for small $\lambda_t$.

If we take the condensate value $m_H = 2m$, then the problem simplifies considerably because the above term is proportional to

$$s - 4m^2 = 4p^2 \qquad (9)$$

where as indicated previously $p$ is the c.m. momentum in $t\bar{t}$ scattering. If we define

$$x = \frac{p^2}{m^2} \qquad (10)$$

and, therefore, $\qquad s = 4m^2(1+x)$

then from (7), (8) and (9), for small $\lambda_t$,

$$N \approx -\frac{\lambda_t^2 x}{2} \qquad (11)$$

$$D \approx x$$

Even though the N/D ratio above is a constant we want to retain the individual x-dependence to examine the possible shifts in x.

We choose the subtraction point for D in the region where D is real, along the negative x-axis. Let this point be $x = -x_0$, then making two subtractions, expression (5) is modified to

$$D(x) = D(-x_0) + (x+x_0)D'(-x_0) - \frac{(x+x_0)^2}{\pi}\int_0^\infty \frac{dx'}{(x'-x)(x'+x_0)^2} \frac{1}{16\pi}\sqrt{\frac{x'}{x'+1}} N(x') \quad (12)$$

Similarly for N(x).

To accomplish (11) we write the first two terms in D as x, and write D (and similarly) N as,

$$D(x) = x - \frac{(x+x_0)^2}{16\pi^2}\int_0^\infty \frac{dx'}{(x'-x)(x'+x_0)^2}\sqrt{\frac{x'}{x'+1}} N(x') \quad (13)$$

$$N(x) = -\frac{\lambda_t^2 x}{2} + \frac{(x+x_0)^2}{16\pi^2}\int_{-\infty}^{x_0} \frac{dx'}{(x'-x)(x'+x_0)^2} D(x') \operatorname{Im} T(x') \quad (14)$$

We note that for small $\lambda_t$ we recover (11) and (7).

We will examine the zeroes of D in the first iteration i.e. we examine the zeroes of (13) by substituting for N the first term on the right hand side of (14). The D-function is then

$$D(x) = x + \frac{(x+x_0)^2}{32\pi^2}\int_0^\infty \frac{dx'\, \lambda_t^2 x'}{(x'-x)(x'+x_0)^2}\sqrt{\frac{x'}{x'+1}} \quad (15)$$

Notice that when $\lambda_t^2$ is very small, then the zero of D(x) is given by x=0 which gives the condensate mass ($m_H = 2m$) for the Higgs boson. However, if $\lambda_t$ is sufficiently large then the zeroes of D are likely to shift.

As for the choice of the subtraction point $x = -x_0$, where $D(x)$ equals the condensate value $x$, we take advantage of the fact that the top color model and condensate models are characterized by a scale. It is, therefore, reasonable to expect that the subtraction point be related to that scale. If we take $\Lambda$ to be the scale parameter in the s-variable i.e. $|s_0| = \Lambda$ then from (10)

$$x_0 \approx \frac{\Lambda^2}{4m^2} \tag{16}$$

We have left $\lambda_t^2$ inside the integral because in actual fact $\lambda_t$ depends on $\mu^2$ as given by (1). That is, it depends on $p^2$ and, therefore, on $x$. Clearly then, depending on the functional form of $\lambda_t^2(\mu)$, the integration in (15) can be very complicated, particularly since (1) predicts a divergent behavior. In fact if we, as emphasized by Bardeen et al [5], impose the composite condition $\lambda_t \to \infty$ as $\mu \to \Lambda$, where $\Lambda$ is the compositeness scale then the solution of (1) turns out to be

$$\lambda_t^2(\mu) = \frac{1}{2b\ln\left(\frac{\Lambda^2}{\mu^2}\right)} \quad , \qquad b > 0 \tag{17}$$

If we insert this expression in (15) then the integral becomes infinite because of the singularity at $\mu \approx \Lambda$.

A more convergent way may be to extrapolate $\lambda_t^2$ at $\mu^2 = \mu_0^2$ linearly through $\mu^2 = \Lambda^2$ by using (1). If we write $\lambda_t^2(\mu_0^2) = \lambda_0^2$ we obtain from (1)

$$\lambda_t^2(\mu) = \lambda_0^2 + 2b\lambda_0^4 \ln\left(\frac{\mu^2}{\mu_0^2}\right) \tag{18}$$

where

$$\lambda_0 = \text{top-Yukawa coupling at the top mass}$$

$$\mu_0 = \text{top mass} = m$$

$$\mu = \sqrt{s}$$

Therefore, for $\lambda_t^2$ inside the integral in (15) we can write

$$\lambda_t^2(\mu) = \lambda_0^2 + 2b\lambda_0^4 \ln[4(1+x)] \tag{19}$$

The integral in (15) is then cut-off at $\mu^2 = \Lambda^2$, or equivalently at $x = x_0$.

The equation (15) can then be written as

$$x = F(x, \lambda_0, \Lambda) \tag{20}$$

where

$$F(x, \lambda_0, \Lambda) = -\frac{(x+x_0)^2}{2} \int_0^{x_0} \frac{dx' \, \alpha_t(x') \, x'}{(x'-x)(x'+x_0)^2} \sqrt{\frac{x'}{x'+1}} \tag{21}$$

where

$$\alpha_t(x) = \alpha_0 + (32\pi^2 b) \, \alpha_0^2 \ln[4(1+x)] \tag{22}$$

$$\alpha_0 = \frac{\lambda_0^2}{16\pi^2} \tag{23}$$

In the following we take $b = \frac{3}{8\pi^2}$ from the MSSM model where there are two Higgs doublets.

From (20) and (21) it is clear that for $\alpha_0 \ll 1$ we recover the condensate result x=0 (i.e. $m_H = 2m$). As $\alpha_0$ is increased, however, the solution of (20), that is the intersection of the straight line given by the left hand side of (20) with $F(x, \lambda_0, \Lambda)$, will give a non-zero value of x. The corresponding Higgs mass is then given by

$$m_H = 2m\sqrt{1+x}. \qquad (24)$$

There are two possible solutions of (20): one corresponds to the region x<0 when both sides of (20) are negative. It involves the low mass region of the Higgs boson. There is also a possible solution for x>0 in a region where $F(x, \lambda_0, \Lambda)$, because it is a principal part integral, becomes positive. This is the second Higgs state. The situation is schematically described in Fig.1.

In Figure 2 we have plotted curves for $m_H$ = 100 and 200 Gev in terms of the two free parameters $\lambda_0$ and $\Lambda$. The dotted line in the figure is the Pagles-Stockar expression [8] which relates $\lambda_0$ to the scale parameter $\Lambda$

$$\frac{\lambda_0^2}{16\pi^2} = \frac{1}{N_c \ln\left(\frac{\Lambda^2}{m^2}\right)} \qquad (25)$$

with $N_c$=3. The intersection of the dotted line with the two curves correspond to the allowed values of $\lambda_0$, and $\Lambda$, for each value of $m_H$.

Only the following narrow corridor in the $(\lambda_0, \Lambda)$ is found to participate in generating Higgs masses between 100 and 200 Gev:

$$3.6 < \lambda_0 < 3.8 \tag{26}$$

$$1.2 \text{ Gev} < \Lambda < 1.6 \text{ Gev}$$

This parameter range is consistent with the top-color models [1], [2], [3].

There will also be another solution of (20) for positive values of $x$, as explained earlier. This corresponds to the second Higgs particle, $m_H^*$. Because the integrand in (21) has an (artificial) infinity at a finite point $x = x_0$ along the integration path, we move the upper limit to $\infty$ as the integral is then convergent. The solution to (20) is found to be

$$m_H^* \approx 1-1.5 \text{ Tev}$$

for the range given by (26), which is, of course, close to the scale parameter.

## Conclusion

We have shown that low mass values for the Higgs boson consistent with the $Z^0$-width precision measurements can be obtained in the condensate models provided that the top-Yukawa coupling is large. The magnitude of the coupling constant at the top mass ($\lambda_0 \approx 3.7$) is consistent with the estimates of top-color models and corresponds to a scale $\Lambda \approx 1.4$ Tev implying that the fermion mass generation has a different scale from the scale for electroweak symmetry breaking. Because the top-Yukawa coupling is large, its evolution at high energies through renormalization group equation plays an important role. Furthermore, in combination with the low mass state a second Higgs around 1 Tev is also generated.

We are grateful to Professor Jose Wudka for several helpful discussions. This work was supported in part by the U.S. Department of Energy under Contract No: DE-F603-94ER40837.

References

[1] C. T. Hill, Phys. Lett. B 266, 419 (1991)

[2] C. T. Hill, Phys. Lett. B 345, 483 (1995); D. Kominis Phys. Lett. B 358, 312 (1995); J.D. Wells, hep-ph/9612292; M. Spira and J.D. Wells, hep-ph/9711410.

[3] R.S. Chivkula, B.A. Dobrescu, H. Georgi and C.T. Hill, hep-ph/9809470.

[4] Y. Nambu, report EFI 88-39 (July1988), published in the proceedings of the *Kazimierz 1988 Conference on New Theories in physics*, ed. T. Eguchi and K. Nishijima; in the proceedings of the *1988 International workshop on New Trends in Strong Coupling Gauge Theories, Nagoya, Japan*, ed. Bando, Muta, and Yamawaki (World Scientific, 1989); report EFI-89-08 (1989); Also see V.A. Miransky, M. Tanabashi and K. Yamawaki, Mod. Phys. Lett. A4 (1989) 1043; Phys. Lett. B 221 (1989) 177; W.J. Marciano, Phys. Lett. 62 (1989) 2793.

[5] W.A. Bardeen, C.T. Hill and M. Lindner, Phys. Rev. D41, 1647 (1990).

[6] G. Giacomelli and R. Giacomelli *Results from High energy $e^+ e^-$ Collisions*, Lecture Notes, DFUB98/23; LEP Electroweak working group. A combination of preliminary LEP electroweak measurements and constraints on the Standard Model, LEPEWWG/98-01 (1998).

[7]  ALEPH Collaboration, P. Teixeira-Dias, *Search for the SM Higgs Boson at the LEP2 collider at $\sqrt{s}=189$ Gev*, CERN-EP/98-144 (1998); DELPHI Collaboration, P. Aberu et al. *A search for neutral Higgs Bosons in the MSSM and in models with two scalar field doublets*, Euro. Phys. Journal C5 (1998) 19; L3 Collaboration, M. Acciarri et al. *Search for the SM Higgs boson in $e^+e^-$ interactions at $\sqrt{s}=189$ Gev*, CERN-EP98/52 (1998); OPAL Collaboration, G. Abbiendi et al. *Search for Higgs bosons in $e^+e^-$ collisions at $\sqrt{s}=189$ Gev*. CERN-EP/98-173 (1998); The LEP working group for Higgs bosons searches, *Lower bounds for the SM Higgs boson mass from combining the results fo the 4 LEP experiments*, CERN-EP/98-145 (1998).

[8] H. Pagels and S. Stockar, Phys. Rev. D20, 2947 ( 1979), also see reference [14].

[9] M.Jacob and G.C. Wick, Ann. Phys. (NY) 7, 404 (1959).

[10] B. Lee and R. Sawyer, Phys. Rev. 127, 2266 (1962); Also see D. Amati, S. Fubini and A. Stanghellini, Phys. Lett. 29 (1962).

[11] R. Blankenbecler and M.L. Goldberger, Phys. Rev. 126, 766 (1962); M.L. Godberger, M.T. Grisaru, S.W. McDowell and D. Wong, Phys.

Rev. 120, 2250 (1960). Also see R.G. Newton, *Scattering Theory of Waves and particles,* 2$^{nd}$ ed. Springer-Verlag, New York, 1982.

[12] G.F. Chew, *S-matrix Theory of Strong Interactions*, Benjamin, New York (1962).

[13] S. Gasiorowitz, *Elementary Particle Physics*, John Wiley & Sons Inc, New York (1966).

[14] T. Gherghetta, Phys. Rev. D50, 5985 (1994).

[15] Y. Nambu and G. Jona-Lasinio, Phys. Rev. 122, 345 (1961); ibid Phys. Rev 124, 246 (1961).

Figure Captions:

Fig. 1. A schematic plot of $x$ and $F(x, \lambda_0, \Lambda)$ as a function of $x$. The intersection points indicate the (bound) state solutions of (20) for a fixed value of $(\lambda_0, \Lambda)$.

Fig. 2. The solutions of (20) corresponding to the Higgs boson mass $m_H = 100$ and $200$ Gev plotted as functions of $\lambda_0$ and $\Lambda$. The dotted curve is the Pagles-Stockar [8] expression (25).

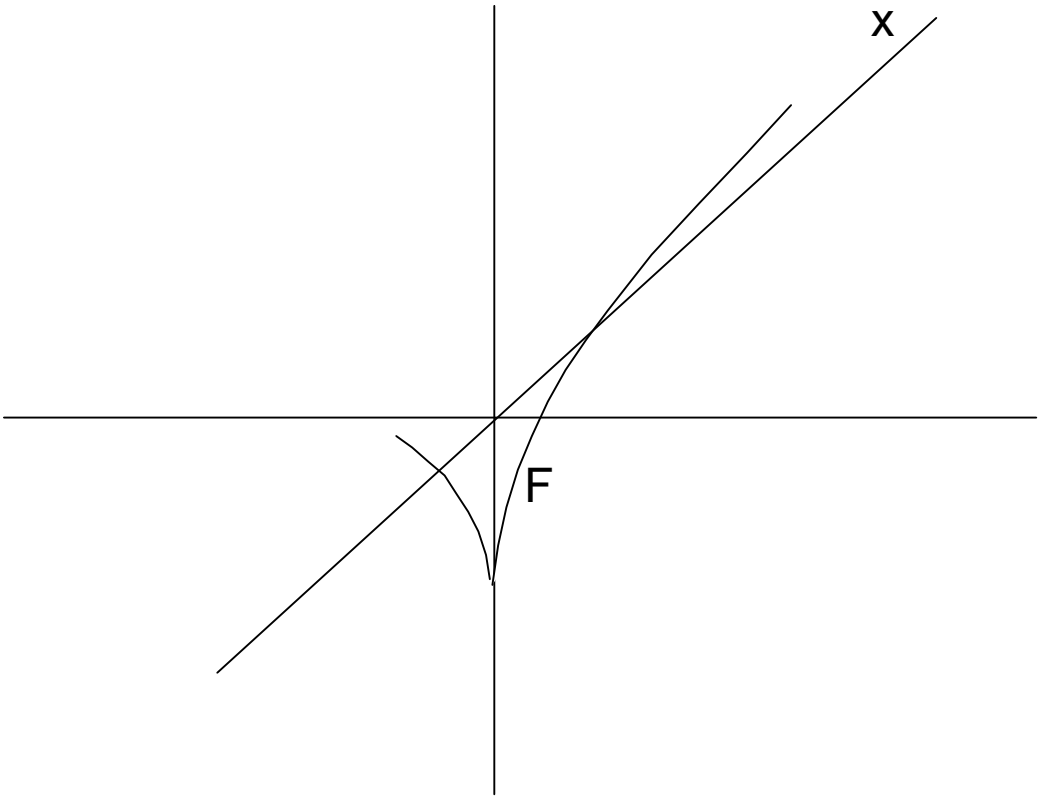

Figure 1.

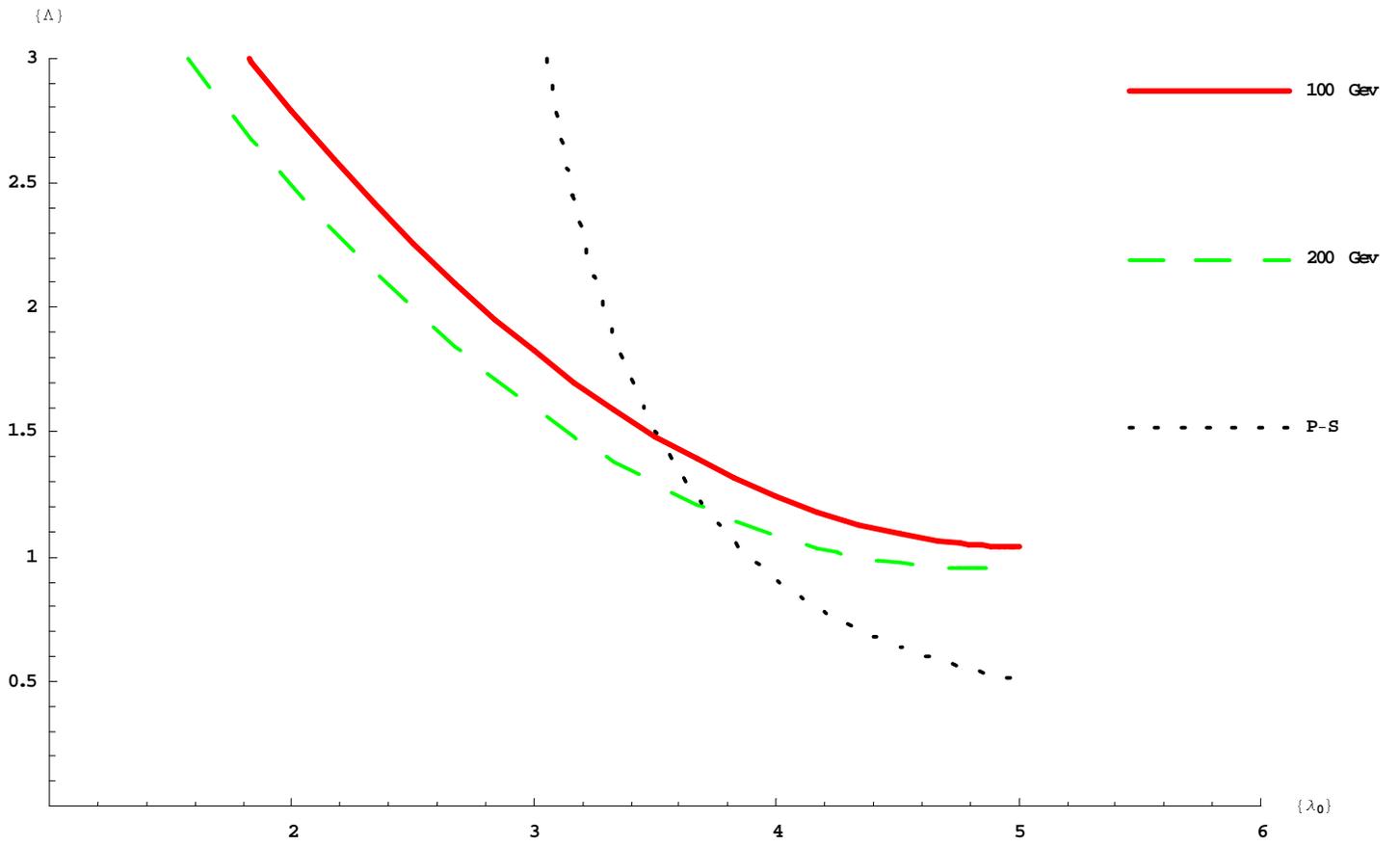

Figure 2.